\documentclass[twocolumn,english]{IEEEtran}
\usepackage[colorlinks,citecolor=blue,urlcolor=blue,bookmarks=false,hypertexnames=true]{hyperref} 
\usepackage{dblfloatfix}
\usepackage{color}
\usepackage{array}
\usepackage{units}
\usepackage{textcomp}
\usepackage{amsmath}
\usepackage{graphicx}
\usepackage{amssymb}
\usepackage{times,amsmath,epsfig}
\usepackage{graphicx}
\usepackage{gensymb}
\usepackage{amsmath}
\usepackage{subfigure}
\usepackage{psfrag}
\usepackage{epsfig}
\usepackage{cite}
\usepackage{multirow}
\usepackage{multicol}
\usepackage{babel}
\usepackage{indentfirst, float}
\graphicspath{ {images/} }
\usepackage{caption}
\usepackage{picinpar}
\usepackage{url}
\usepackage{flushend}
\usepackage{colortbl}
\usepackage{soul}
\usepackage{pifont}
\usepackage{alltt}

\title{Use Cases for Time-Frequency Image Representations and Deep Learning Techniques for Improved Signal Classification}

\author{ \href{https://orcid.org/0000-0003-0276-9289}{\includegraphics[scale=0.06]{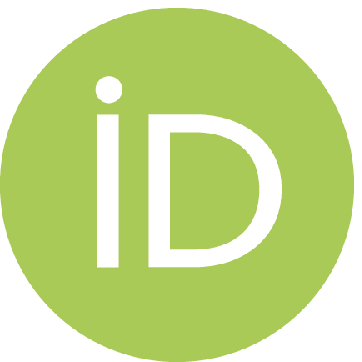}\hspace{1mm}Mehmet Parlak}
\thanks{The authors are at the Department of Electrical and Electronics Engineering, Ozyegin University, Istanbul (e-mail: mehmet.parlak@ozyegin.edu.tr)} \\
	Department of Electrical and Electronics \\
	Ozyegin University\\
	Cekmekoy, Istanbul 34794 \\
	\texttt{mehmet.parlak@ozyegin.edu.tr} \\
}

\hypersetup{
pdftitle={A template for the arxiv style},
pdfsubject={q-bio.NC, q-bio.QM},
pdfauthor={Mehmet Parlak},
pdfkeywords={First keyword, Second keyword, More},
}

\begin{document}
\maketitle

\begin{abstract}
 Time-frequency images (TFIs) provide a joint time-frequency representation of a signal and have become an effective tool for analyzing, characterizing, and processing non-stationary signals. Deep learning (DL) techniques have become versatile for signal classification, enabling the automatic extraction of relevant features from raw data. In this paper, we present two use cases on the time-frequency transformation and deep learning techniques for signal classification, where signals are first pre-processed and transformed into TFIs, and their features are then extracted through deep learning neural networks and classification algorithms. The specific methods and algorithms used may vary depending on the particular application, therefore different methods for creating TFIs; the Short-Time Fourier Transform (STFT), Fourier-based Synchrosqueezing Transform (FSST), Wigner Ville distribution (WVD), Smoothed Pseudo-Wigner distribution (SPWD), Choi-Williams distribution (CWD), and Continuous Wavelet Transform (CWT) are investigated. The performance of various deep learning, and convolutional neural network (CNN) models such as  ResNet-50, ShuffleNet, and Squeezenet are evaluated for their accuracy of classification in different applications and the results are compared with the results of the conventional machine learning and ensemble methods such as Multilayer Perceptrons (MLP), Support Vector Machine (SVM), Random Forest (RF), Decision Tree (DT), and XGboost. The results of this research demonstrate that significant improvements in signal classification accuracy can be achieved by leveraging the combined power of TFIs, and deep learning models. These advances have found practical applications in a wide range of fields, including radar signal classification, stability analysis of power systems, speech and music recognition, and biomedical signal characterization.

\end{abstract}


\section{Introduction}
Time-frequency transformations (TFTs) are mathematical algorithms that convert time-domain signals into time-frequency images (TFIs), also known as spectrograms. In other words, TFIs are the visual output of TFTs, and graphical representations of how the energy or power of a signal is distributed across both time and frequency domains, providing a detailed analysis of the signal's characteristics \cite{cohen}. They are used to identify the temporal and spectral properties and extract features from signals that are not easily visible in the time or frequency domain alone. 

The time-frequency analysis pipeline for signal classification involves signal pre-processing, obtaining time-frequency images, feature extraction, and classification using specific methods and algorithms depending on the application. Once the TFI of the signal has been obtained, it can be provided as input into the neural networks so that classifications and predictions can be made based on the raw data. Hence, a signal classification problem is translated into an image classification.

\section{Applications}
\label{sec:headings}
The most common use of TFIs is in the analysis and processing of non-stationary signals, i.e., signals that change over time, mostly encountered in fields such as communication and radar signals classification, transient stability analysis (TSA), human activity and gesture recognition, fault detection, audio and speech recognition, image processing, biomedical signal processing, vibration analysis, and financial forecasting.

\begin{figure*}[htbp!]
	\centering
	\noindent \includegraphics[width=1\textwidth]{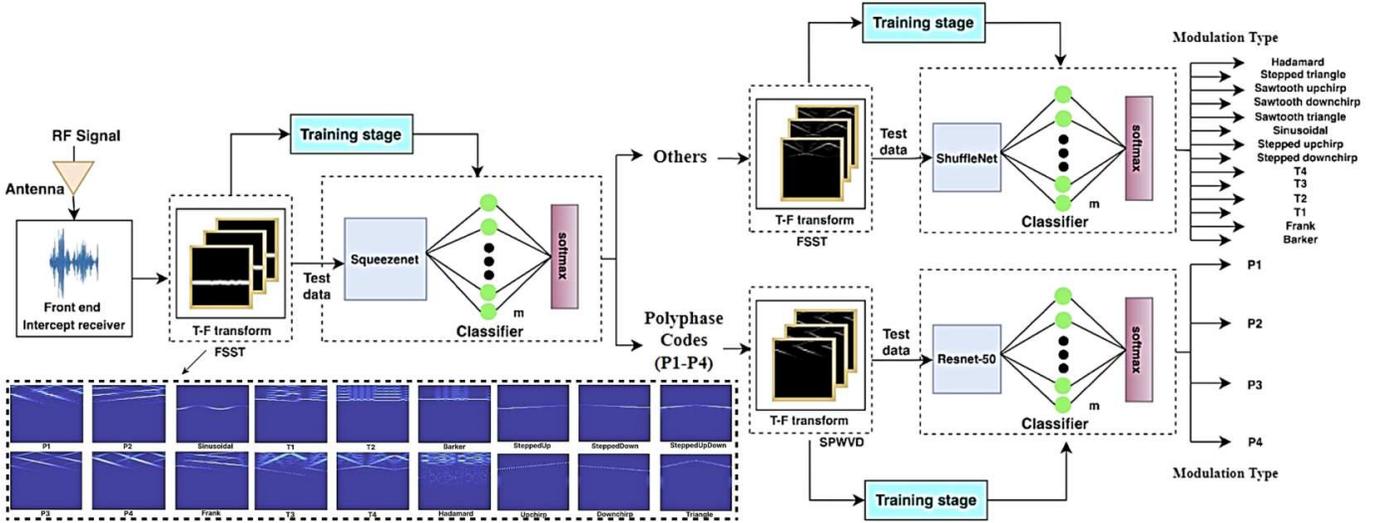}\\
	\caption {Proposed multi-stage architecture and method for radar signal classification.}
	\label{fig:Rradar}
\end{figure*}

\subsection{TFI for Radar Signal Classification}
Radar signals are used to detect and track objects in the environment, such as aircraft, ships, or ground vehicles. These signals are typically complex and can contain information about the distance, speed, and direction of the target, as well as the target's physical characteristics. However, radar signals can also be affected by noise, interference, and other sources of distortion, which can make it difficult to extract useful information from the signal. 

Some radar waveforms may be better suited for detecting certain types of targets or in certain environmental conditions, while others may be less effective \cite{Haykin06}. Therefore, cognitive radar systems use multiple types of waveforms to detect and classify targets. With an increase in the number of different types of waveforms utilized in cognitive radar applications, the individual performance of classification accuracy for each waveform tends to vary even more due to their distinct characteristics such as their frequency, pulse duration, modulation, and polarization. Furthermore, the classification accuracy of each waveform can be influenced by factors such as the target's shape, size, and composition, the presence of clutter or interference, the range and angle of the target, and interactions of the waveform with the environment and targets. This variability in performance can make it challenging to optimize the cognitive radar system and can impose stringent requirements on its associated circuits, such as baseband analog correlation circuits \cite{Parlak11}, \cite{Li13}, \cite{Parlak12}, \cite{Li15}. It also requires careful selection and coordination of the different radar waveforms used.

As depicted in Fig. \ref{fig:Rradar}, the architecture for radar waveform classification employs several modular blocks of TFTs with varying strengths and CNN networks optimized for each stage based on computational requirements and accuracy to classify eighteen different types of radar waveforms \cite{Guven22}. By utilizing various TFT methods and CNN networks the overall system can adapt to a broader range of waveform types at various stages. Fig. \ref{fig:Rradar} illustrates the process of transforming eighteen distinct signal waveforms into high-resolution TFIs using FSST in the initial stage, followed by classification as either "polyphase coded (P1-P4)" or "other" using a binary classifier, where Squeezenet is used as the binary classification model due to its quick and robust performance. For signals classified as P1-P4, the system proceeds to further classify them using SPWVD, since WVD is better suited for PM signals, and ResNet-50 is employed as the classifier. If the output of the first stage is classified as "other," then the signals are transformed using FSST, and a fourteen-class classifier with the Shufflenet, a competent model to extract line-shaped features from FM signals even at low SNR levels to facilitate classification. 

\begin{figure}[htbp!]
	\centering
	\noindent \includegraphics[width=0.48 \textwidth]{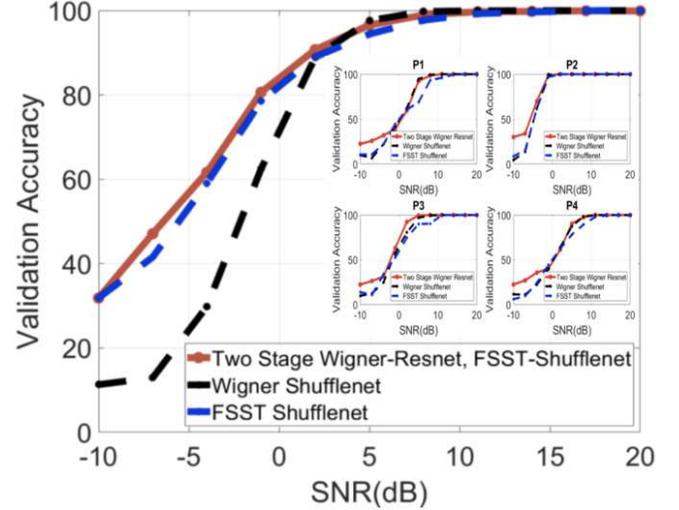}\\
	\caption {Accuracy comparison for overall average classification of eighteen waveforms, and P1-P4 radar waveforms}
	\label{fig:radar_results}
\end{figure}

To compare our proposed method, we tested two 18 classes classifiers with two different transformations: SPWVD and FSST. The comparison of the three methods is shown in Fig. \ref{fig:radar_results}, which demonstrates that the average classification accuracy for all eighteen waveforms is slightly enhanced at low SNR levels through the use of multi-stage modular TFIs and CNNs. When examining the individual performances of each waveform, there are noticeable differences. As our base system employs FSST, the utilization of SPWVD for polyphase-coded signals has resulted in improved performance. Fig. \ref{fig:radar_results} demonstrates that the performance of different polyphase-coded waveforms P1-P4 is improved at low SNR values, although the proposed system performs similarly to the traditional Wigner-Shufflenet system at high SNR values. 

\begin{figure*}[htbp]
	\centering
	\noindent \includegraphics[width=1\textwidth]{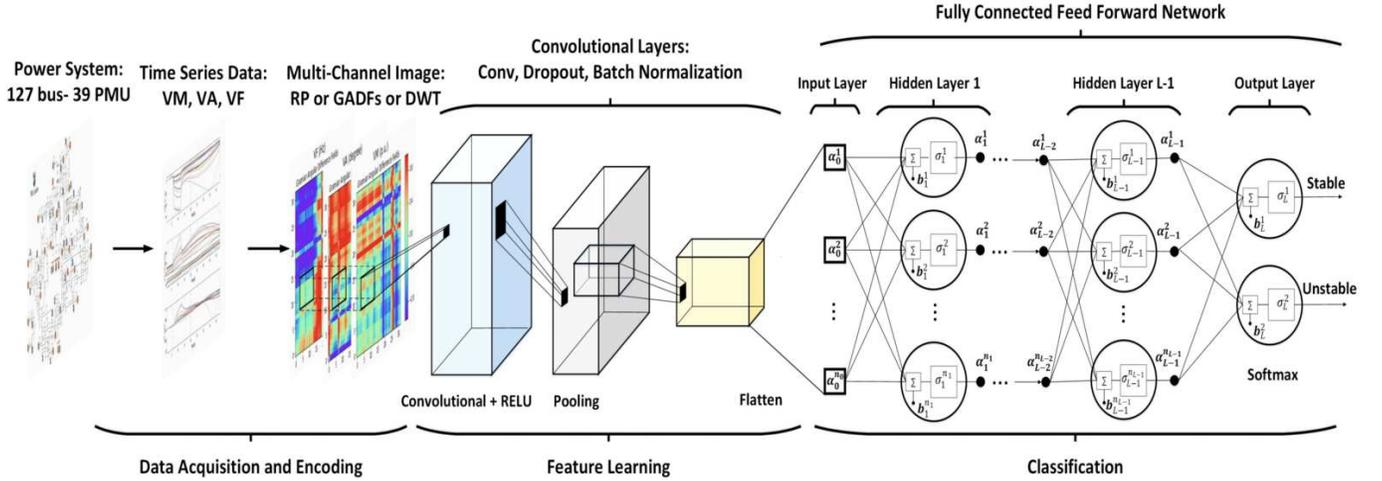}\\
	\caption {The proposed architecture and method for power transient stability analysis (TSA).}
	\label{fig:powertsa}
\end{figure*}

\vspace{-4mm}
\subsection{TFIs for Transient Stability of Power Systems}

In the context of power systems, a robust, fast, and real-time transient stability analysis (TSA) is essential for ensuring reliable, efficient, and safe operation. TFIs can provide a high-resolution time-frequency representation of the power system signals and their evolution over time, which reveals the dynamic behavior of the system during a sudden disturbance, such as a fault or a generator trip. Such a disturbance can cause transient instability and lead to a cascading failure of the system, which can result in blackouts and damage to equipment. Therefore, it is crucial to detect and mitigate these events as quickly as possible through TFIs, predicting the stability of the system and detecting any anomalies. Moreover, TFIs can also aid in condition monitoring and predictive maintenance of power systems. By continuously monitoring the system signals and analyzing the TFIs, we can detect any gradual changes in the frequency content, which may indicate the presence of faults or degradation in the system components. This information can be used to schedule maintenance activities and prevent equipment failure, leading to reduced downtime and cost savings.

\begin{figure}[htbp!]
	\centering
	\noindent \includegraphics[width=0.48 \textwidth]{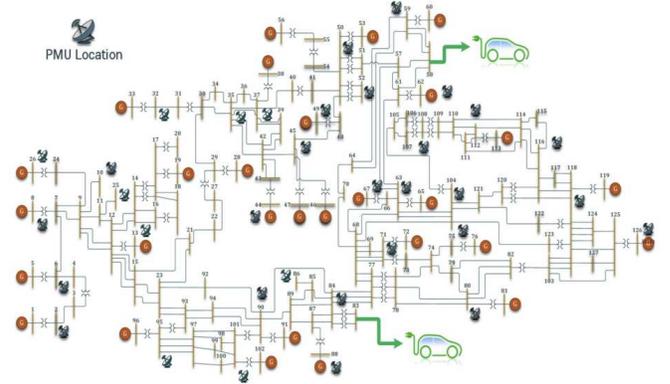}\\
	\caption {The WSCC 127-bus system's single line diagram showcasing the placement of both the EVFCS and PMUs}
	\label{fig:GridPMU}
\end{figure}

Fig. \ref{fig:powertsa} presents a framework of TFIs and 2D-CNNs to monitor the transient stability before and after electric vehicle fast-charging stations (EVFCS) are loaded into the smart power grid \cite{Behdadnia21}. The phasor measurement units (PMUs) placed in the WSCC 127-bus system, shown in Fig. \ref{fig:GridPMU}, provides a vast collection of synthetic data  to train our ML/DL models. Certain types of electrical quantities in the sequential PMU data are utilized to forecast the system stability. In this case, the voltage magnitude $V_{M}$ is chosen as the input feature to train conventional ML models such as XGboost, RF, DT, MLP.

On the other hand, to train the CNN, a 3-channel input, TFIs of voltage magnitude $V_{M}$, voltage angle $V_{A}$, voltage frequency $V_{F}$, is created with a time window of 0.1s through RP, GADF, and DWT transformations. Then, the TFIs are fed into 2D-CNNs. During the feature learning stage, a series of operations are performed, including convolution (filtering), activation, and pooling. Trainable kernels are used to convolve the multi-channel input, and the resulting output is passed through an activation function. The data is then subjected to a pooling operation, which reduces its resolution and makes it more resistant to small variations in previously learned features. Finally, at the end of this stage, the feature maps are flattened for further processing. After flattening the processed data into one-dimensional representations, it is fed into a fully connected feed-forward network for the final classification stage. In the last layer of the network, a softmax function is utilized as the activation function to produce the final prediction result. This function is highly appropriate for binary classification problems, including the prediction of power system transient stability, as it is capable of transforming a vector of real values into probabilities that add up to 1.

The PMU data captured in the WSCC 127-bus system were used to evaluate the performance of different classifiers, based on three metrics. While ACC represents the overall accuracy,  the TUR and TSR represent the proportion of correctly predicted unstable instances, and the proportion of correctly predicted stable instances, respectively. As shown in Table  I, the DWT and CNN combination provides the highest accuracy, and reduces the computational cost of training and inference as the DWT can reduce the size of the input data by decomposing the images into frequency sub-bands, which can be processed more efficiently by the CNN.

\vspace{3.5mm}
\noindent TABLE I Accuracy results before/after integration of EVFCS.
\vspace{-5mm}
\begin{figure}[htbp!]
	\centering
	\noindent \includegraphics[width=0.48 \textwidth]{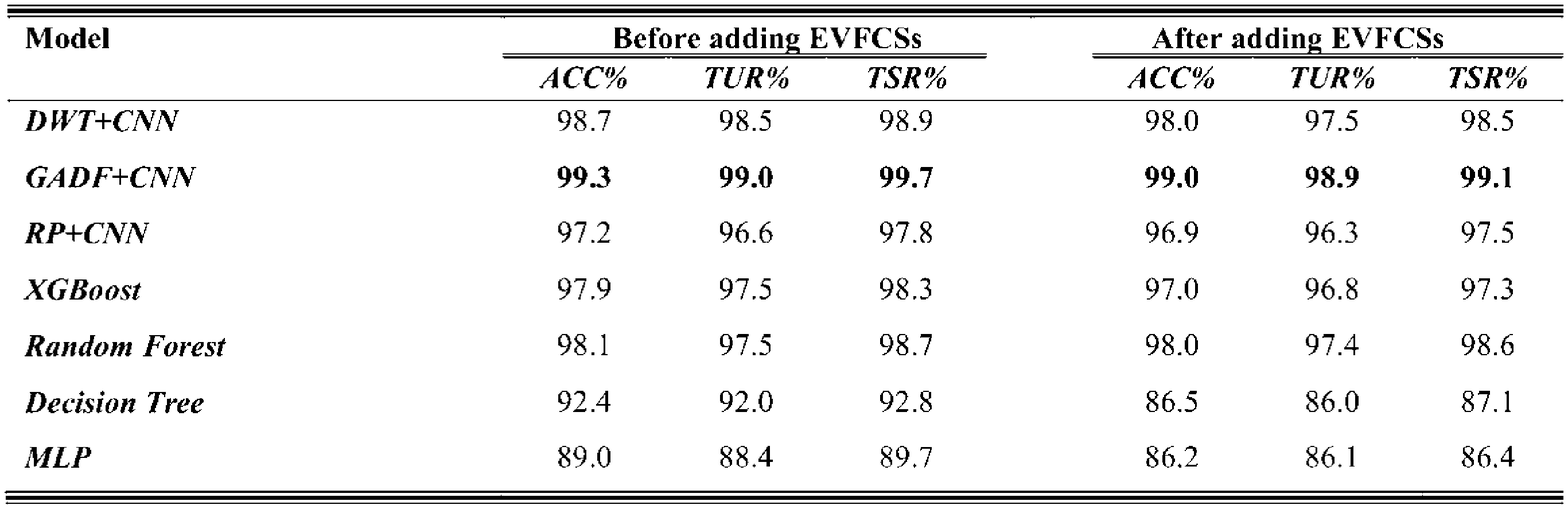}\\
	\label{fig:tablep}
\end{figure}

\section{Conclusion}

The time-frequency analysis pipeline for signal classification involves signal pre-processing, obtaining TFIs, feature extraction, and classification using specific methods and algorithms depending on the application. This research shows that combining TFIs, DL models and algorithms improves signal classification accuracy in fields like radar signal classification and power systems transient stability analysis. The choice of appropriate TFTs and DL models, algorithms, and methods for each stage is highly dependent on the specific application and the nature of the signal being analyzed. 

 In the first part of the paper, the focus is on deep learning several TFIs of various radar waveform types accurately, as recent advancements in deep learning have revolutionized the field of radar signal processing with highly effective and efficient solutions \cite{Geng21}. In this context, the results presented in this research shows that different TFTs have varying strengths, and utilizing an architecture with modular TFTs and DL models show great promise in improving individual radar signal classification accuracy while being highly scalable and adaptable. The FSST provides focused representations of radar signals, but the Fourier and wavelet-based methods have limited resolution. This limitation is overcome by WVD, which uses autocorrelation to improve resolution. Compared to other members of Cohen's class of TFDs, WVD distributes signal energy more effectively over a joint time-frequency domain. Therefore, WVD is preferred separating out P1-P4 since identifying the structures of polyphase signals is challenging without high-resolution time-frequency images.

In the second part of the paper, the potential benefits of combining time-frequency representations and deep learning models for the transient stability analysis of EV-integrated smart power grid (before and after EVFCSs are loaded in) is demonstrated. In such modern power systems, the timely implementation of control measures to prevent cascading failures and blackouts is crucial, and this requires accurate and rapid prediction of transient stability status. The proposed approach utilizes a hybrid-type simulator to generate a large and realistic set of synthetic PMU data in a feasible amount of time. The hybrid-type simulation involves the detailed simulation of PMUs and EV loads in the electromagnetic transient type (EMT) domain. During the learning phase, raw PMU data $V_{M}$, $V_{A}$, $V_{F}$ serves as predictor variables. By employing a combination of time series imaging techniques such as RP, GADF, and DWT, and utilizing 2D-CNN, informative spatial-temporal features are extracted from the raw PMU data. The incorporation of these features lead to a significant improvement in accuracy performance of the TSA compared to other ML methods.

Overall, this research contributes to the growing body of knowledge on signal processing in the time-frequency domain and highlights the potential of deep learning models in improving signal classification accuracy. The findings presented in this paper is relevant to researchers and practitioners working in the field of signal processing and related areas, providing insights into the benefits of combining time-frequency analysis and deep learning techniques for diverse applications. While the combination of time-frequency images and deep learning models has shown promising results, some challenges such as determining the appropriate time-frequency transformation and optimal deep learning architecture automatically, and avoiding overfitting due to the high dimensionality of time-frequency images require careful pre-processing, regularization, and hyperparameter tuning.

\section{Future Work}
In the context of machine learning, pre-processing refers to the transformation of raw input data into a form that is more amenable to analysis by a model. As future work, we are specifically interested in pre-processing the TFI, which is the key input to neural networks. The main challenge with the TFI is that it often contains noise and other artifacts that can negatively impact the performance of downstream tasks. This is where advanced pre-processing layers come in. Therefore, the development of advanced pre-processing layers aimed at either denoising or accentuating basic shapes in the TFI will be prioritized. 

These layers will leverage state-of-the-art techniques such as Hough transformation \cite{Hassanein15} and denoising neural networks such as residual dense neural networks (RDNs). Hough transformation is a popular technique in computer vision used for detecting simple shapes, such as lines or circles, in an image. By applying this technique as a pre-processing step, we aim to accentuate these basic shapes in the TFI, which can improve the accuracy and robustness of downstream tasks that rely on shape information. On the other hand, RDNs have shown to be effective in image denoising tasks, as they can effectively model intricate features and textures in images, and learn complex mappings between noisy and clean images \cite{Ramos21}. This makes them well-suited for handling challenging denoising tasks and producing high-quality denoised images even in the presence of complex textures.  By integrating both Hough transformation and RDNs into a new pre-processing layer, we aim to enhance the quality and usefulness of the TFI for downstream tasks.

\bibliography{references}

\begin{thebibliography}{10}
\providecommand{\url}[1]{#1}
\csname url@samestyle\endcsname
\providecommand{\newblock}{\relax}
\providecommand{\bibinfo}[2]{#2}
\providecommand{\BIBentrySTDinterwordspacing}{\spaceskip=0pt\relax}
\providecommand{\BIBentryALTinterwordstretchfactor}{4}
\providecommand{\BIBentryALTinterwordspacing}{\spaceskip=\fontdimen2\font plus
\BIBentryALTinterwordstretchfactor\fontdimen3\font minus
  \fontdimen4\font\relax}
\providecommand{\BIBforeignlanguage}[2]{{%
\expandafter\ifx\csname l@#1\endcsname\relax
\typeout{** WARNING: IEEEtran.bst: No hyphenation pattern has been}%
\typeout{** loaded for the language `#1'. Using the pattern for}%
\typeout{** the default language instead.}%
\else
\language=\csname l@#1\endcsname
\fi
#2}}
\providecommand{\BIBdecl}{\relax}
\BIBdecl

\bibitem{cohen}
L.~Cohen, \emph{{T}ime-{F}requency {A}nalysis}.\hskip 1em plus 0.5em minus
  0.4em\relax Prentice {H}all New Jersey, 1995.

\bibitem{Haykin06}
S.~Haykin, ``{C}ognitive {R}adar: a {w}ay of the {f}uture,'' \emph{IEEE Signal
  Processing Magazine}, vol.~23, no.~1, pp. 30--40, 2006.

\bibitem{Parlak11}
M.~Parlak, J.~F. Buckwalter, and M.~Matsuo, ``{B}idirectional {C}ircuitry for
  {M}illimeter-wave {P}ulse {C}ompression {R}adar,'' in \emph{2011 IEEE
  RadarCon (RADAR)}, 2011, pp. 1062--1067.

\bibitem{Li13}
J.~Li, H.~Mukai, M.~Parlak, M.~Matsuo, and J.~F. Buckwalter, ``{A} 1{G}b/s
  {R}econfigurable {P}ulse {C}ompression {R}adar {S}ignal {P}rocessor in 90nm
  {CMOS},'' in \emph{Proceedings of the IEEE 2013 Custom Integrated Circuits
  Conference}, 2013, pp. 1--4.

\bibitem{Parlak12}
M.~Parlak, M.~Matsuo, and J.~F. Buckwalter, ``{A}nalog {S}ignal {P}rocessing
  for {P}ulse {C}ompression {R}adar in 90-nm {CMOS},'' \emph{IEEE Transactions
  on Microwave Theory and Techniques}, vol.~60, no.~12, pp. 3810--3822, 2012.

\bibitem{Li15}
J.~Li, M.~Parlak, H.~Mukai, M.~Matsuo, and J.~F. Buckwalter, ``{A}
  {R}econfigurable 50-{M}b/s-1 {G}b/s {P}ulse {C}ompression {R}adar {S}ignal
  {P}rocessor with {O}ffset {C}alibration in 90-nm {CMOS},'' \emph{IEEE
  Transactions on Microwave Theory and Techniques}, vol.~63, no.~1, pp.
  266--278, 2015.

\bibitem{Guven22}
I.~Guven, C.~Yagmur, B.~Karadas, and M.~Parlak, ``Classifying {LPI} {R}adar
  {W}aveforms with {T}ime-{F}requency {T}ransformations using {M}ulti-{S}tage
  {CNN} {S}ystem,'' in \emph{2021 IEEE Radio Frequency Integrated 2022 23rd
  International Radar Symposium (IRS)}, 2022, pp. 501--506.

\bibitem{Behdadnia21}
T.~Behdadnia and M.~Parlak, ``{EV}-{I}ntegrated {P}ower {S}ystem {T}ransient
  {S}tability {P}rediction {B}ased on {I}maging {T}ime {S}eries and {D}eep
  {N}eural {N}etwork,'' in \emph{2021 IEEE International Intelligent
  Transportation Systems Conference (ITSC)}, 2021, pp. 3198--3203.

\bibitem{Geng21}
Z.~Geng, H.~Yan, J.~Zhang, and D.~Zhu, ``{D}eep-{L}earning for {R}adar: {A}
  {S}urvey,'' \emph{IEEE Access}, vol.~9, pp. 141\,800--141\,818, 2021.

\bibitem{Hassanein15}
A.~S. Hassanein, S.~Mohammad, M.~Sameer, and M.~E. Ragab, ``{A} {S}urvey on
  {H}ough {T}ransform, {T}heory, {T}echniques and {A}pplications,'' \emph{arXiv
  preprint arXiv:1502.02160}, 2015.

\bibitem{Ramos21}
J.~Gurrola-Ramos, O.~Dalmau, and T.~E. Alarcón, ``{A} {R}esidual {D}ense
  {U}-net {N}eural {N}etwork for {I}mage {D}enoising,'' \emph{IEEE Access},
  vol.~9, pp. 31\,742--31\,754, 2021.

\end{thebibliography}
\bibliographystyle{IEEEtran}
\end{document}